\documentclass[preprint,showpacs,preprintnumbers,amsmath,amssymb]{revtex4}

\usepackage{graphicx}% Include figure files
\usepackage{dcolumn}% Align table columns on decimal point
\usepackage{bm}% bold math

\begin{document}

%\preprint{APS/123-QED}

\title{Non-Linear Deformations of Liquid-Liquid Interfaces Induced by Electromagnetic Radiation Pressure }% Force line breaks with \\

\author{Aslak Hallanger\footnote{Email: aslakhallanger@hotmail.com}, Iver Brevik\footnote{Email: iver.h.brevik@ntnu.no},
Skjalg Haaland\footnote{Email: skjalg.haaland@ntnu.no} \\}

\affiliation{ Department of Energy and Process Engineering,
Norwegian University of Science and Technology, N-7491 Trondheim,
Norway
}%

\author{Roger Sollie\footnote{Email: rsol@statoil.com}}
\affiliation{ Department of Physics, Norwegian University of
Science and Technology, N-7491 Trondheim, Norway, \\
and \\
Statoil Research Centre, N-7005 Trondheim, Norway,
\\
}%
Revised version
\date{ \today}% It is always \today, today,

\begin{abstract}
The idea of working with a near-critical phase-separated liquid
mixture whereby the surface tension becomes weak, has recently
made the field of laser manipulation of liquid interfaces a much
more convenient tool in practice. The deformation of interfaces
may become as large as several tenths of micrometers, even with
the use of conventional laser power. This circumstance
necessitates the use of nonlinear geometrical theory for the
description of surface deformations. The present paper works out
such a theory, for the surface deformation under conditions of
axial symmetry and stationarity. Good agreement is found with the
experimental results of Casner and Delville [A. Casner and J. P.
Delville, Phys. Rev. Lett. {\bf 87}, 054503 (2001); Opt. Lett.
{\bf 26}, 1418 (2001); Phys. Rev. Lett. {\bf 90}, 144503 (2003)],
in the case of moderate power or a broad laser beam. In the case
of large power and a narrow beam, corresponding to surface
deformations of about 50 micrometers or higher, the theory is
found to over-predict  the deformation. Possible explanations of
this discrepancy are discussed.
\end{abstract}

\pacs{42.25.Gy, 42.50.Vk, 82.70.Kj}
\maketitle

\section{\label{sec:level1}Introduction}

The formation, deformation, and breakup of fluid interfaces are
ubiquitous phenomena in nature \cite{moseler00}. One special group
of effects which implies the so-called finite time singularity
\cite{eggers97}, has as its most common example the breakup of
pendant drops driven by gravity \cite{cohen99,chen02}. If external
fields such as electric or magnetic fields are present, one has in
addition extra control parameters. Thus we may mention that
interface instabilities driven by electric fields
\cite{taylor69,oddershede00} are important for many processes such
as electro-spraying \cite{ganan97}, ink-jet printing
\cite{badie97}, or surface-relief patterning \cite{schaffer00}. A
uniform magnetic field can also be useful, for instance for the
purpose of of forming elongated magnetic droplets \cite{bacri83}.
These deformations, as well as those induced by the acoustic
radiation pressure on liquid surfaces \cite{hertz39,elrod89}, have
been used to explore the mechanical properties of fluid interfaces
in a non-contact way \cite{cinbis93,flament96}.

It is noteworthy that laser-induced deformations of the interfaces
of soft materials have not received the same amount of interest in
the past. Most attention has been given to test-particle global
effects such as optical levitation and trapping - cf., for
instance, Refs.~\cite{ashkin97,gussgard92}. The reason for this
circumstance is simple: deformations of fluid interfaces by
optical radiation are ordinarily {\it weak}. For instance, in the
classic experiment of Ashkin and Dziedzic \cite{ashkin73}, a
pulsed single transverse mode doubled Nd:YAG laser (20 pulses per
second) was focused vertically from above on a water surface. The
wavelength of the incident wave was $\lambda_0=0.53 \; \mu$m, the
peak power was $P_{max}=3$ kW (low enough to make nonlinear
effects negligible), and the duration of each pulse was 60 ns. The
beam radius at the waist was reported to be very small,
$\omega_0=2.1\; \mu$m, but has most likely have been somewhat
larger, $\omega_0=4.5\;\mu$m; cf. the dicussion in
Ref.~\cite{brevik79}. The rise of the water surface was small,
about $0.9\; \mu$m, occurring at $t \approx 450 $ ns after the
onset of the pulse. The physical reason for this small surface
elevation is  evidently the large surface tension $\sigma$ =0.073
N/m between air and water. [The theory of the Ashkin-Dziedzic
experiment was worked out in Refs.~{\cite{lai76,brevik79}.] If we
go to the more recent experiment of Sakai {\it et al.}
\cite{sakai01}, we will find that the surface displacement was
even more minute. In this case the displacement was induced by a
continuous wave (cw) Ar$^+$ pump laser (wavelength in vacuum
$\lambda_0=0.514\; \mu$m, maximum power $P_{max}$=0.5 W), and was
probed with a He-Ne laser. For a beam power $P$= 0.3 W and a beam
waist $\omega_0=142\: \mu$m the elevation of the surface was found
to be extremely small, about  2 nm.

Generally speaking it is of considerable interest to be able to
probe the displacement of fluid interfaces in a way that is
non-contact, i.e., which avoids a direct touch of the fluid by
mechanical devices. The only acting force on the fluid surface is
thus the radiation force. As discussed in Ref.~\cite{casner01},
for instance, this kind of force can measure locally the
micromechanical properties of soft biological systems because
artificial membranes \cite{barziv98} or cells \cite{guck00} can be
highly deformable. Sizable effects of the radiation pressure
should thus be measured, facilitating the characterization of
surface elasticity properties.  A major step forward
 was to recognize that one can reduce the
surface tension considerably by working with a phase-separated
liquid mixture, close to the critical point. In this way "giant"
deformations can be achieved. Recent experiments of Casner and
Delville have shown that the displacements can in this way reach
several tenths of micrometers
\cite{casner01a,casner01,casner01b,casner03,casner03a} (cf. also
the recent review \cite{delville04}). Also, there are seen in the
experiments rather remarkable asymmetries with respect to  the
upward/downward direction of the laser beam \cite{casner03a}.

The giant deformations make it necessary to use nonlinear theory
for the description of the surface deflection. The purpose of the
present paper is to present such a calculation, for the typical
case where the elevation is axially symmetric (a condition almost
always satisfied in practice), and in addition stationary.
Comparison with the mentioned experiments will be made. We shall
moreover assume that the laser beam is incident from below only.
The mentioned up/down asymmetry will thus not be treated.

For completeness we emphasize that we are considering nonlinear
theory only in a geometrical sense, in order to describe the
interface bulge. The electromagnetic theory as such is kept on the
conventional  linear level, as we are only considering moderate
laser intensities. The theory for {\it linear} deformations has
been worked out before \cite{lai76,brevik79,casner01,casner01b}.
The analytic solution for the deflection of the surface is
reproduced in Eq.~(\ref{47}).

\section{Derivation of the Governing Equation}

\subsection{Basic electromagnetic theory}

We begin by writing down the  expression for the electromagnetic
volume force density ${\bf f}$ in an isotropic, nonconducting and
nonmagnetic medium \cite{stratton41,brevik79}:
\begin{equation}
{\bf f}=-\frac{1}{2}E^2{\bf \nabla} \varepsilon +\frac{1}{2}\nabla
\left[E^2\rho \left(\frac{\partial \varepsilon}{\partial
\rho}\right)_S \right]
+\frac{\kappa-1}{c^2}\frac{\partial}{\partial t}({\bf E \times
H}). \label{1}
\end{equation}
Here $\bf {E}$ and $\bf{H}$ are the electric and magnetic fields,
$\rho$ is the mass density of the medium (fluid), $\varepsilon$ is
the permittivity, and $\kappa =\varepsilon/\varepsilon_0$ is the
relative permittivity where  $\varepsilon_0$ denotes the
permittivity of vacuum.

Let us comment on the various terms in Eq.~(\ref{1}), beginning
with the last term. This term is called the Abraham term, since it
follows from Abraham's electromagnetic energy-momentum tensor. The
term is experimentally detectable under special circumstances at
low frequencies \cite{brevik79}, but not at optical frequencies,
at least not under usual stationary conditions. The Abraham term
simply fluctuates out.

The middle term in Eq.~(\ref{1}) is the electrostriction term.
When seen from within the optically denser medium (the medium with
the highest $n$), the electrostriction force is always
compressive. Whether this kind of force is detectable in a static
or a stationary case, depends on whether the experiment is able to
measure local pressure distributions within the compressed region
or not. Moreover, in a dynamic case the velocity of {\it sound} is
an important factor. If the elastic pressure in the fluid has
sufficient time to build up, then the electrostriction force will
not be detectable when measuring the gross behavior of a fluid
such as the elevation of its surface. Such is usually the case in
optics. The time required for the counterbalance to take place, is
of the same order of magnitude as the time needed for sound waves
to traverse the cross section of the laser beam. For a beam width
around $10\; \mu $m, this yields a time scale for counterbalance
of the order of 10 ns. For instance,  in the Ashkin-Dziedzic
experiment \cite{ashkin73}  a detailed calculation verifies this
time scale; cf. Fig.~9 in \cite{brevik79}.

Another point worth mentioning in connection with the
electrostriction term is that that we have written $(\partial
\varepsilon/\partial \rho)_S$ as an adiabatic partial derivative.
This seems most natural in optical problems in view of the rapid
variations of the field, at least in connection with laser pulses.
In many cases it is however legitimate to assume that the medium
is nonpolar, so that we need not distinguish between adiabatic and
isothermal derivatives. The permittivity depends on the mass
density only. Then derivative can be written simply as
$d\varepsilon /d\rho$, and is calculable from the
Clausius-Mossotti relation. In this way we can write Eq.~(\ref{1})
in the following form, when omitting the last term,
\begin{equation}
{\bf f}=-\frac{1}{2}\varepsilon_0E^2{\bf \nabla}\kappa +
\frac{1}{6} \varepsilon_0{\bf \nabla}
\left[E^2(\kappa-1)(\kappa+2)\right]. \label{2}
\end{equation}

Finally, we have the first term in Eq.~(\ref{1}), which may be
called the Abraham-Minkowski force, since it follows equally well
from the Abraham and the Minkowski energy-momentum tensors:
\begin{equation}
{\bf f}^{AM}=-\frac{1}{2}\varepsilon_0 E^2{\bf \nabla}\kappa.
\label{3}
\end{equation}
This is the only term that we have to take into account in
practice in optics, under usual circumstances. We see that this
force is equal to zero in the homogeneous interior of the medium,
and acts in the inhomogeneous boundary region only. By integrating
the normal component of the Abraham-Minkowski force density across
the boundary, we obtain the surface force density which can
alternatively be evaluated as the jump of the normal component of
the electromagnetic Maxwell stress tensor.

In the following we  focus the attention on the force term in
Eq.~(\ref{3}) only.

\subsection{Surface tension and radiation forces on a curved
surface}

 Let us assume that there is established a stationary curved surface $z=h(x,y)$
 distinguishing
 two fluids, a lower fluid (1) and an upper fluid (2), the
 equilibrium position being determined by the balance of gravity, surface
 tension, and radiation pressure.
The undisturbed  position of the surface is the $xy$ plane.
Because of the surface tension coefficient $\sigma$, there will
  be a normal stress proportional to the mean curvature of the
  surface:
  \begin{equation}
  p_2-p_1=\sigma \left(\frac{1}{R_1}+\frac{1}{R_2}\right),
  \label{4}
  \end{equation}
  $R_1$ and $R_2$ being the principal radii of curvature at the
  surface
  point considered. If $R_1$ and $R_2$ are positive, $p_2-p_1 >0$.
  This means that the pressure is greater in the medium whose
  surface is convex. It is  useful  to have in mind  the following general formula
  for the mean curvature $1/R_1+1/R_2$:
  \begin{equation}
  \frac{1}{R_1}+\frac{1}{R_2}=\frac{h_{xx}(1+h_y^2)+h_{yy}(1+h_x^2)-2h_{xy}h_xh_y}
  {(1+h_x^2+h_y^2)^{3/2}},
  \label{5}
  \end{equation}
  where $h_x=\partial h/\partial x$, etc. Our convention is such that the curvature is
   positive if the surface is concave upwards. Also, we note that the
  unit normal vector $\bf n$ to the surface is
  \begin{equation}
  {\bf n}=(1+h_x^2+h_y^2)^{-1/2}(-h_x,-h_y,1). \label{6}
  \end{equation}
  The normal points upwards, from medium 1 to medium 2.

  Assume now that there is a monochromatic electromagnetic wave
  with electric field vector ${\bf E}^{(i)} ({\bf r})e^{-i\omega
  t}$ incident from below, in the positive $z$ direction. The
  direction of the incident wave vector ${\bf k}_i$ is thus given
  by the unit vector
  \begin{equation}
  \hat{{\bf k}}_i=(0,0,1)
  \label{7}
  \end{equation}
in medium 1. When this wave impinges upon the surface, it becomes
separated into a transmitted wave ${\bf E}^{(t)}$ and a reflected
wave ${\bf E}^{(r)}$, propagating in the directions of $\hat{{\bf
k}}_t$ and $\hat{ {\bf k}}_r$, respectively. We assume, in
conformity with usual practice,  that the  waves can locally be
regarded as plane waves and that the surface can locally be
regarded as plane.  The plane of incidence is formed by the
vectors ${\hat {\bf k}}_i$ and $\bf n$; we call  the angle of
incidence $\theta_i$ and the angle of transmission  $\theta_t$.
See Fig.~1.  Moreover, we let ${\bf E}_\parallel$ and ${\bf
E}_\perp$ be the components of $\bf E$ parallel and perpendicular
to the plane of incidence, respectively. The expressions for the
energy flux transmission coefficients $T_\parallel$ and $T_\perp$
for a plane wave incident upon a  boundary surface are (cf.
\cite{stratton41}, p. 496):
\begin{equation}
T_\parallel=\frac{n_2}{n_1}\frac{\cos \theta_t}{\cos\theta_i}
\left(\frac{E_\parallel^{(t)}}{E_\parallel^{(i)}}\right)^2=
\frac{\sin 2\theta_i \sin
2\theta_t}{\sin^2(\theta_i+\theta_t)\cos^2(\theta_i-\theta_t)},
\label{8}
\end{equation}
\begin{equation}
T_\perp=\frac{n_2}{n_1}\frac{\cos \theta_t}{\cos \theta_i}
\left(\frac{E_\perp^{(t)}}{E_\perp^{(i)}}\right)^2= \frac{\sin
2\theta_i \sin 2\theta_t}{\sin^2(\theta_i+\theta_t)}. \label{9}
\end{equation}
When dealing with an unpolarized radiation field, one usually
averages over the two polarizations and represents the
transmission coefficient by the single entity
\begin{equation}
\langle T\rangle=\frac{1}{2}(T_\parallel +T_\perp).\label{10}
\end{equation}
Consider now the electromagnetic surface force density, which we
will call $\bf{\Pi}$. As mentioned above, $\bf{\Pi}$ can be found
by integrating the normal component of the volume force density
across the surface boundary layer. From Eq.~(\ref{3}) it follows
that the surface force acts normal to the surface, and that it is
directed towards the optically thinner medium.

We introduce the intensity $I$ of the incident beam,
\begin{equation}
I=\varepsilon n_1 c \langle {E^{(i)}}^2\rangle \label{11}
\end{equation}
(in the case of azimuthal symmetry $I=I(r)$), and let $\alpha$
denote the angle between ${\bf E}^{(i)}$ and the plane of
incidence,
\begin{equation}
E_\parallel^{(i)}= E^{(i)}\cos \alpha, \quad
E_\perp^{(i)}=E^{(i)}\sin \alpha . \label{12}
\end{equation}
Then, we can write the surface force density as
\begin{equation}
{\bf \Pi}=-\frac{I}{2c}\frac{n_2^2-n_1^2}{n_2}\frac{\cos
\theta_i}{\cos
\theta_t}\left[(\sin^2\theta_i+\cos^2\theta_t)T_\parallel\cos^2\alpha+T_\perp
\sin^2\alpha \right]\bf{n}. \label{13}
\end{equation}
When ${\bf E}^{(i)}={\bf E}_\parallel^{(i)}$ or ${\bf
E}^{(i)}={\bf E}_\perp^{(i)}$ (i.e., $\alpha =0$ or $\pi/2$) it is
often convenient to express $\bf{\Pi}$ as
\begin{equation}
{\bf{\Pi}}=\frac{n_1 I}{c}\cos^2\theta_i \left(1+R- \frac{\tan
\theta_i}{\tan \theta_t}T \right)\bf{n}, \label{14}
\end{equation}
where $R=1-T$ is the reflection coefficient. This expression has
been derived before \cite{borzdov93,casner03,casner03a}. It holds
also in the hydrodynamic nonlinear case.
 In connection with the mentioned Bordeaux experiments
 \cite{casner01,casner01a,casner01b,casner03,casner03a}
, the upper liquid was always the optically denser one. Thus
$n_2>n_1$,
 the direction of $\bf \Pi$ is antiparallel to $\bf{n}$, and the
 force acts downward, normal to the surface.

The case of normal incidence yields
\begin{equation}
T_\parallel=T_\perp=\frac{4n_1n_2}{(n_2+n_1)^2}, \label{15}
\end{equation}
\begin{equation}
{\bf\Pi}=-\frac{2n_1I}{c}\frac{n_2-n_1}{n_2+n_1}{\bf n}.
\label{16}
\end{equation}

\subsection{ Cylindrical symmetry}

We  henceforth assume cylindrical symmetry, using standard
cylinder coordinates $(r,\theta, z)$. There is no variation in the
azimuthal direction, $\partial h/\partial \theta=0$. With the
notation $h_r=\partial h/\partial r$ we have
\begin{equation}
\cos \theta_i=\frac{1}{\sqrt{1+h_r^2}},\quad \sin
\theta_i=\frac{h_r}{\sqrt{1+h_r^2}}. \label{17}
\end{equation}
Together with analogous expressions for $\theta_t$ this can be
inserted into Eq.~(\ref{13}) to yield
\begin{equation}
{\bf \Pi}=-\frac{2n_1I(r)}{c}\frac{1-a}{1+a}f(h_r, \alpha){\bf n},
\label{18}
\end{equation}
where $a$ is the relative refractive index,
\begin{equation}
a=n_1/n_2 <1, \label{19}
\end{equation}
and $f(h_r,\alpha)$ is the function
\[
f(h_r,\alpha)=\frac{(1+a)^2}{\left[a+\sqrt{1+(1-a^2)h_r^2}\right]^2}
\]
\begin{equation}
\times
\left\{\sin^2\alpha+\frac{1+(3-a^2)h_r^2+(2-a^2)h_r^4}{\left[ah_r^2+\sqrt{1+(1-a^2)h_r^2}\right]^2}\,\cos^2\alpha
\right\}. \label{20}
\end{equation}
When the surface is horizontal, $h_r=0$, we have $f=1$, and $\bf
\Pi$ reduces to the expression (\ref{16}).

A peculiar property of the expression (\ref{20}) facilitating
practical calculations is that it is quite insensitive with
respect to variations in the polarization angle $\alpha$,
especially in the case when $a$ is close to unity, which is in
practice most important.  Thus if we draw curves for
$\Pi(\theta_i)$ versus $\theta_i$ for various input values of
$\alpha$ in the whole region $0<\alpha <90^o$ (not shown here), we
will find that the curves lie close to each other. For practical
calculations involving unpolarized light it is thus legitimate to
replace $f(h_r,\alpha)$ by its average with respect to $\alpha$.
As $\langle \sin^2\alpha \rangle=\langle \cos^2
\alpha\rangle=1/2$, we can then write the surface force density as
\begin{equation}
{\bf \Pi}=-\frac{2n_1I(r)}{c}\frac{1-a}{1+a}f(h_r){\bf n},
\label{21}
\end{equation}
where $f(h_r)$ is equal to $f(\alpha, h_r)$ averaged over
$\alpha$,
\begin{equation}
f(h_r)=(1+a)^2\frac{1+(2-a^2)h_r^2+ah_r^2\sqrt{1+(1-a^2)h_r^2}+h_r^4}
{\left[a+\sqrt{1+(1-a^2)h_r^2}\right]^2
\left[ah_r^2+\sqrt{1+(1-a^2)h_r^2}\right]^2}. \label{22}
\end{equation}

This expression is valid also in the case of hydrodynamic
nonlinearity. Note again that $\bf \Pi$ is the normally-directed
force per unit area of the oblique {\it liquid surface}.

Finally, let us consider the force balance for the liquid column,
assuming stationary conditions. When $n_2>n_1$ the surface tension
force which acts upward, has to balance the combined effect of
gravity and electromagnetic surface force, which both act
downward. When the surface is given as $h=h(r,\theta)$, the mean
curvature can be written as
\begin{equation}
\frac{1}{R_1}+\frac{1}{R_2}=\frac{1}{r}\frac{\partial}{\partial
r}\frac{rh_r}{\sqrt{1+h_r^2+(h_\theta/r)^2}}
+\frac{1}{r^2}\frac{\partial}{\partial
\theta}\frac{h_\theta}{\sqrt{1+h_r^2+(h_\theta/r)^2}}, \label{23}
\end{equation}
 with sign conventions the same as in Eq.~(\ref{5}). Thus for
 azimuthal symmetry,
 \begin{equation}
 \frac{1}{R_1}+\frac{1}{R_2}=\frac{1}{r}\frac{d}{dr}\frac{rh_r}{\sqrt{1+h_r^2}},
 \label{24}
 \end{equation}
 and the force balance becomes \cite{mitani02,casner03}
 \begin{equation}
 (\rho_1-\rho_2)gh(r)-\frac{\sigma}{r}\frac{d}{dr}\left[\frac{rh_r}{\sqrt{1+h_r^2}}\right]=\Pi(r).
 \label{25}
 \end{equation}
 This equation follows from considering the equilibrium of a
 liquid column having unit base area. Here $\Pi(r)$ is the
 pressure scalar, i.e., ${\bf \Pi}(r)=\Pi(r){\bf n}$. Thus
 $\Pi(r)<0$.

 What expression to insert for $\Pi(r)$ in Eq.~(\ref{25}), depends on
 the physical circumstances. Thus in the case of an unpolarized
 laser beam, we may use either the expression (\ref{14}) with
 $R=\langle R\rangle,\; T=\langle T\rangle$, or alternatively use
 the expression (\ref{21}). We will follow the latter option here. As  noted, there is no
 restriction imposed on the magnitude of the slope of the surface.

 \section{Solution of the Nonlinear Equation}

 It is advantageous to introduce nondimensional variables.
  Let us first define the capillary length
 $l_C$ and the Bond number $B$,
 \begin{equation}
 l_C=\sqrt{\frac{\sigma}{(\rho_1-\rho_2)g}},\quad
 B=\left(\frac{\omega_0}{l_C}\right)^2, \label{26}
 \end{equation}
 $\omega_0$ being the radius of the beam waist. The Bond number describes the  strength
 of buoyancy relative to the Laplace force.  If $B\ll 1$, gravity is  much weaker than the Laplace force.
 (The Bordeaux experiments covered the region $10^{-3}<B<4$.)
  We then define the nondimensional radius $R$ and the nondimensional
  height $H(R)$ as
 \begin{equation}
 R=\frac{r}{\omega_0}, \quad H(R)=\frac{h(r)}{l_C}. \label{27}
 \end{equation}
 The fact that in practice $a=n_1/n_2$ is very close to one, makes
it at first sight possible to  simplify the right hand side of the
governing equation (\ref{25}).  Namely, from Eq.~(\ref{22}) one
would expect that $f(h_r) \rightarrow 1$. However, the situation
is more delicate due to nonlinearity: if we keep $f(h_r)$ in the
formalism and calculate the elevation $h(r)$, we will find that
$f$ gets a pronounced dip within the region where the beam is
strong. Typically, if we draw a curve for $f=f(R)$ versus the
nondimensional radius $R$, we will see that $f$ drops from 1 to
about 0.3 when $R$ lies about 0.5. Mathematically, this is because
the high steepness of the surface makes $h_r$ (or $H_R$) large
enough to influence $f$ significantly in a narrow region even when
$a$ is close to unity. Assuming a Gaussian incident beam,
 \begin{equation}
 I(r)=\frac{2P}{\pi \omega_0^2}\, e^{-2r^2/\omega_0^2}, \label{28}
 \end{equation}
 with $P$ the beam power,
 we may write the governing equation (\ref{25}) as a nonlinear
 differential equation for $H$:
 \begin{equation}
 BH-\frac{H_{RR}+\frac{1}{R}H_R+\frac{1}{BR}H_R^3}{\left(1+\frac{1}{B}H_R^2\right)^{3/2}}=-F
 e^{-2R^2}f(H_R). \label{29}
 \end{equation}
 Here $F$ is a positive constant at fixed temperature,
 \begin{equation}
 F=\frac{2(n_2-n_1)P}{\pi cg (\rho_1-\rho_2)l_C^3}, \label{30}
 \end{equation}
which can for practical purposes be written as, since $(\partial
n/\partial \rho)_T=-1.22\times 10^{-4}\;{\rm m^3/kg}$,
\begin{equation}
F=\frac{2}{\pi cg}\left( -\frac{\partial n}{\partial
\rho}\right)_T\frac{P}{l_C^3}=26400 \frac{P}{l_C^3}. \label{31}
\end{equation}
In the last equation, the dimension of $l_C$ is $\mu$m.

 %The boundary conditions are
 %\begin{equation}
 %H_R\Big|_{R=0}=0, \quad H(R\rightarrow \infty)=0. \label{32}
 %\end{equation}
 The two quantities $l_C$ and $a$ will vary with the temperature $T$
 in accordance with the theory of critical phenomena. Thus for the
 density contrast $\Delta \rho=\rho_1-\rho_2$ we have
 \begin{equation}
\Delta \rho=(\Delta \rho)_0\left( \frac{T-T_C}{T_C}\right)^\beta,
\label{32}
\end{equation}
where $\beta=0.325,\; (\Delta \rho)_0=285$ kg/$\rm{m}^3$,
$T_C=$308.15 K being the critical temperature above which  the
mixture separates into two different phases. Similarly
\begin{equation}
\sigma=\sigma_0\left( \frac{T-T_C}{T_C}\right)^{2\nu}, \label{33}
\end{equation}
with $\nu=0.63,\; \sigma_0=1.04\times 10^{-4}$ N/m.  More details
can be found in Refs.~\cite{casner01,casner01b}. We give here the
practically useful formulas for $a$ and $l_C$:
\begin{equation}
a=1-0.0238 \left( \frac{T-T_C}{T_C}\right)^{0.325}, \label{34}
\end{equation}
\begin{equation}
l_C=193\left(\frac{T-T_C}{T_C}\right)^{0.468} \quad (\mu {\rm m}).
\label{35}
\end{equation}

These two quantities are the only parameters that vary with
temperature. There are thus  three parameters in all in the
problem, namely $T$, the beam power $P$, and the beam waist
$\omega_0$. Nondimensionally, the last two parameters correspond
to $F$ and $B$, Eqs.~(\ref{31}) and (\ref{26}).

\subsection{Numerical solution}

It is convenient to let $H$ be positive downwards, so that in the
formalism below we will replace $H$ with $-H$.

We  start from the nondimensional governing equation in the form
\begin{equation}
\frac{1}{R}\frac{d}{dR}\left(
\frac{RH_R}{\sqrt{1+H_R^2/B}}\right)-BH=-Fe^{-2R^2}f(H_R),
\label{36}
\end{equation}
with boundary conditions
\begin{equation}
H_R(0)=0, \quad H(\infty)=0. \label{37}
\end{equation}
We  use a two-point method to solve the nonlinear differential
equation iteratively. Define
\begin{equation}
K=\sqrt{1+H_R^2/B}, \quad S=Fe^{-2R^2}f(H_R), \label{38}
\end{equation}
and let $G=H_R/K$. We obtain the following first order system:
\begin{equation}
\frac{dH}{dR}=K G, \label{39}
\end{equation}
\begin{equation}
\frac{dG}{dR}+\frac{G}{R}-BH=-S, \label{40}
\end{equation}
with  boundary conditions
\begin{equation}
G(0)=0, \quad H(\infty)=0. \label{41}
\end{equation}
We linearize the equations by means of lagging, i.e., we use
values for $H_R$ from the last iteration in the nonlinear
functions $K$ and $S$. Using a nonuniform grid with $n$ grid
points, we integrate the equations between two grid points $j$ and
$j+1$, letting $m$ be the midpoint and $\Delta R_j$ the distance
between the points. We obtain
\begin{equation}
H_{j+1}-H_j=K_m \frac{\Delta R_j}{2}(G_j+G_{j+1}), \label{42}
\end{equation}
\begin{equation}
G_{j+1}-G_j+\frac{1}{R_m}\frac{\Delta
R_j}{2}(G_j+G_{j+1})-B\frac{\Delta R_j}{2}(H_j+H_{j+1})=-\Delta
R_j \,S_m. \label{43}
\end{equation}
Here $H_R$ in $K_m$ and $S_m$ are evaluated as
\begin{equation}
H_R=\frac{\bar{H}_{j+1}-\bar{H}_j}{\Delta R_j}, \label{44}
\end{equation}
where the $\bar{H}$'s are values from the previous iteration. With
$n$ grid points there are $n-1$ intervals and $n-1$ sets of
equations. This confirms with the fact that there are $2n$
quantities $H$ and $G$; since there are  two boundary conditions
there remain $2n-2$ equations in all.

To start the iterations we give initial values for $H=const.\,
e^{-R^2}$. To deal with the boundary condition at infinity, we
need in practice to replace "infinity" with a finite upper limit
$R=R_e$. The solution falls off quite slowly with $R$, so to use
the naive condition $H(R_e)=0$ would require $R_e$ to be very
large. To avoid calculating the long tail of the solution, we can
find a better boundary condition by using the lowest order term in
an asymptotic expansion for $H$. When $R$ is large, $S \sim
e^{-2R^2}$, and $H_R^2$ is very small so that $S \approx 0$,
$\sqrt{1+H_R^2/B}\approx 1$. Equation (\ref{40}) becomes
\begin{equation}
\frac{1}{R}\frac{d}{dR}\left(RH_R\right)-BH=0. \label{45}
\end{equation}
To lowest order this equation has the asymptotic solution $H\sim
R^{-1/2}e^{-\sqrt{B}R}$, which in turn implies that
\begin{equation}
G=H_R=-\left( \frac{1}{2R}+\sqrt{B}\right)H. \label{46}
\end{equation}
We take this condition to replace the condition $H=0$ at $R=R_e$.

We solved the discretized equations using a
Block-Bidiagonal-Matrix-Algorithm, developed by one of the authors
(S.H.). Our programming language was MATLAB.

\subsection{Results}

First, the following question naturally arises: at which powers
$P$ will the nonlinear correction begin to be important? And what
magnitudes of the centerline deformation ($r=0$) does this
correspond to? To get insight into this issue we constructed a
number of figures (not shown here) for the surface height $h(r)$
versus $r$, for various temperature differences $T-T_C$ and beam
radii $\omega_0$, and for various values of $P$. For each
parameter set we made the calculation in two ways, viz. when
taking the nonlinear correction into account, and when omitting
it. Of course, there is a transitional region  and no sharp limit
distinguishing the linear and nonlinear regions. Our conclusions,
based upon visual inspection of the curves, were that under normal
conditions the linear region can be taken to prevail until $P
\approx 200$ mW. When $P > 300$ mW, nonlinear effects turn up.
Generally, the nonlinear deformations are higher than the linear
ones. (To give an example: at $P=220$ mW the nonlinear centerline
deformation was found to be only 2\% higher than the linear
deformation, whereas at $P=330$ mW it was 5\% higher .)

From Refs.~\cite{casner01,casner01b} we recall that in the linear
regime we have the following simple formula for the centerline
height $h^{lin}(r)$ (here in physical variables):
\begin{equation}
h^{lin}(r)= \frac{P}{2\pi cg}\left(\frac{\partial n}{\partial
\rho}\right)_T\int_0^\infty
\frac{k\,dk}{1+k^2l_C^2}\,e^{-k^2\omega_0^2/8}\,J_0(kr).
\label{47}
\end{equation}
In the following we show three figures, each of them corresponding
to given values of $T-T_C$ and $\omega_0$. Each figure is based
upon a  full nonlinear calculation. First, Fig.~2 shows how $h(r)$
varies with $r$ when $T-T_C=2.5$ K and $\omega_0=4.8$ $\mu$m.
According to Eqs.~(\ref{35}), (\ref{26}) this corresponds to
$l_C$=20.3 $\mu$m, $B$=0.0560. The powers are $P=\{300, 600,
1200\}$ mW. As mentioned above, we had to replace "infinity" with
a finite outer limit $R_e$. Numerical trials observing  the
sensitivity of calculated centerline deformations showed that the
choice $R_e=9$ was large enough. (For instance, an increase of
$R_e$ from 9 to 10 would lead only to minute differences, the
first three digits in the centerline deformation being the same.)
Because of the cylindrical symmetry, only one half of the
displacements ($r>0$ in the figure) need to be shown. It is seen
that both powers 600 mW and 1200 mW lead to deformations much
greater than 20 $\mu$m, and are  clearly in the nonlinear region.
The theoretical deflections for the three given values of the
power are read off from the figure to be $\{12, 56, 112\}$ $\mu$m,
respectively.

Our choice of input parameters makes the figure directly
comparable to Fig.~6.1 in Casner's thesis \cite{casner01a}. The
experimental centerline displacements estimated from the photos
are about 10 $\mu$m for $P=300$ mW, 45 $\mu$m for $P=600$ mW, and
 75 $\mu$m for $P=1200$ mW. The theoretical predictions are thus
in this case  larger than the observed ones, especially for the
highest value of $P$.  Moreover, one difference which is most
noticeable is the absence of the observed "shoulder" in the
theoretical solution in the case of large $P$. The shoulder occurs
experimentally when the laser illumination is from below. There is
at present no theory  capable of describing this phenomenon.
Mathematically, the establishment of the shoulder seems to be
related to an instability; the real deflection jumps from one
class of solutions of the nonlinear differential equation to
another class. Video records actually show "jumps" in the surface
when it gets formed, thus supporting our conjecture about an
instability phenomenon. As for the observed {\it width} of the
surface displacement, there is good agreement with the theoretical
prediction.

Fig.~3 shows analogous results for the case $T-T_C=2.5$ K,
$\omega_0=8.9$ $\mu$m, thus a considerably broader beam. Here
$l_C=$20.3 $\mu$m, $B=0.193$. The powers are $P=\{ 360, 600,
890\}$ mW. In this case, the value $R_e=7$ was found to be
sufficient, for the same reasons as above.  Our results  can be
compared with Fig.~6.3 in Casner's thesis \cite{casner01a}. The
theoretical centerline displacements for the three mentioned cases
of $P$ are $\{ 10, 19, 47\}$ $\mu$m, which all agree well with the
observed values. Also in this case there occurs a shoulder
experimentally, but it is not so pronounced as in the previous
case.

Finally, in Fig.~4 we show the case $T-T_C=3$ K, $\omega_0=5.3$
$\mu$m, corresponding to $l_C$=22.1 $\mu$m, $B= 0.0576$, for
powers $P=\{ 300, 590, 830\}$ mW. Again, the outer nondimensional
radius $R_e=9$ was found to be appropriate. The theoretical
centerline deflections are now seen from the figure to be
$\{10,39,65\}$ $\mu$m, respectively, for the given values of $P$.
We may compare this with the photos shown in Fig.~2 in
Ref.~\cite{casner03a} or Fig.~VI.5 in \cite{delville04}: the
corresponding experimental dispacements are about $\{10,40,55\}$
$\mu$m. Also this time we see that the agreement between theory
and experiment is good for low powers, but that the theory
over-predicts the displacement when the power gets large. The
last-mentioned effect is generally most pronounced when the radius
of the laser beam  is small.

\section{Summary, and final remarks}

The "giant" deformations of fluid interfaces recently obtained in
the experiments of Casner and Delville
\cite{casner01,casner01a,casner01b,casner03,casner03a,delville04}
with the use of moderate laser beam powers $P$ ($P$ typically
lying between 500 and 1000 mW) have accentuated the need of taking
into account nonlinear geometrical effects in the theoretical
description of the interface deformation. As a rule of thumb,
inferred from a visual inspection of the figures, nonlinear
effects are expected to be appreciable when the deformations
become larger than about 15 $\mu$m. When the radius $\omega_0$ of
the laser beam is small, typically $\omega \sim 5$ $\mu$m, a power
$P \sim $ 1 watt may easily lead to deflections around 100 $\mu$m.
The physical reason for the occurrence of giant deformations is
the lowering of surface tension caused by working with a
phase-separated liquid  mixture close to the critical point.

The nonlinear governing equation in nondimensional form can be
taken as in Eq.~(\ref{29}) or, what was found more convenient, as
in Eq.~(\ref{36}) where $G=H_R/K$ is considered as the dependent
variable. We solved the set of equations (\ref{39})-(\ref{41})
numerically. As a consistency check, we made also an analogous
calculation starting from Eq.~(\ref{29}), and got the same
results. Figures 2-4 show some examples of our calculations; these
are all directly comparable with the Casner-Delville experiment.

Some general conclusions that can be made from our calculations,
are the following:

1. For given values of $T-T_C$ and $\omega_0$, a larger $P$ causes
a larger deformation.

2.  For a given $T-T_C$, a smaller $\omega_0$ causes a larger and
narrower deformation.

3.  For a given $\omega_0$, a smaller $T-T_C$ causes a larger and
narrower deformation.

4.  Very large beam waists ($\omega_0 \sim 20-30$ $\mu$m) are not
able to cause a nonlinear deformation, not even for the largest
$P$ and smallest $T-T_C$ investigated in the Casner-Delville
experiment.

5.  For small $T-T_C$ and small $\omega_0$ (for instance $T-T_C
=2.5$ K and $\omega_0 = 4.5$ $\mu$m), a power of 300 mW is not
enough to cause a nonlinear deformation. However, a further
decrease in temperature, such as to the value $T-T_C$=1.5 K, will
take also the 300 mW-induced deformation into the nonlinear
regime.

All the items listed above are expected on physical grounds. A
large incident power concentrated on a narrow cross section means
a large electromagnetic field intensity and thus a large surface
force. The enhanced deformation for small $T-T_C$ is due to the
fact that the restoring buoyancy force ($\sim \Delta \rho$) and
Laplace force ($\sim \sigma$) vanish at $T=T_C\,$;  cf.
Eqs.~(\ref{32}) and (\ref{33}).

Concretely, when comparing our results with the Casner-Delville
observations, we find that for broad beams the agreement between
theory and experiment is quite good; cf. our discussion of Fig.~3
above. There is however a considerable theoretical over-prediction
of the deflection in the case of narrow beams and high powers.
Most strikingly, this is shown in the case $\omega_0$=4.8 $\mu$m,
$P$=1200 mW, as discussed in connection with Fig.~2. The physical
reason for this discrepancy is not known. It may be related to the
production of heat in the liquid in the presence of the strong
field, or to the loss of radiation energy because of scattering
from the non-avoidable corrugations on the liquid interface.
Perhaps the most intriguing possibility is that the discrepancy is
related to the reflection of radiation energy from the interface
"shoulder", which is seen to be produced in strong fields when the
illumination is from below. This effect is most likely related to
an instability; the system decides to switch from one class of
solutions of the nonlinear governing equation to another class. To
our knowledge, no explanation exists of this effect.

%\section*{Figure captions}
-
%\noindent FIG.1  Definition sketch of the displaced surface. The
%laser illumination is from below.

%\noindent FIG.2  Theoretical height $h(r)$ of displaced surface
%versus radius $r$ when $T-T_C=2.5$ K, $\omega_0=4.8$ $\mu$m, for
%three different laser powers $P$. The undisturbed surface is at
%$h(r)=0$.

%\noindent FIG.3  Same as Fig.~2, but with $T-T_C=2.5$ K,
%$\omega_0=8.9$ $\mu$m.

%\noindent FIG.4  Same as Fig.~2, but with $T-T_C=3$ K,
%$\omega_0=5.3$ $\mu$m.

\eject
\section*{Figures}

\vskip 1.0cm
\begin{figure}[h]
\centering
\includegraphics[height=8cm]{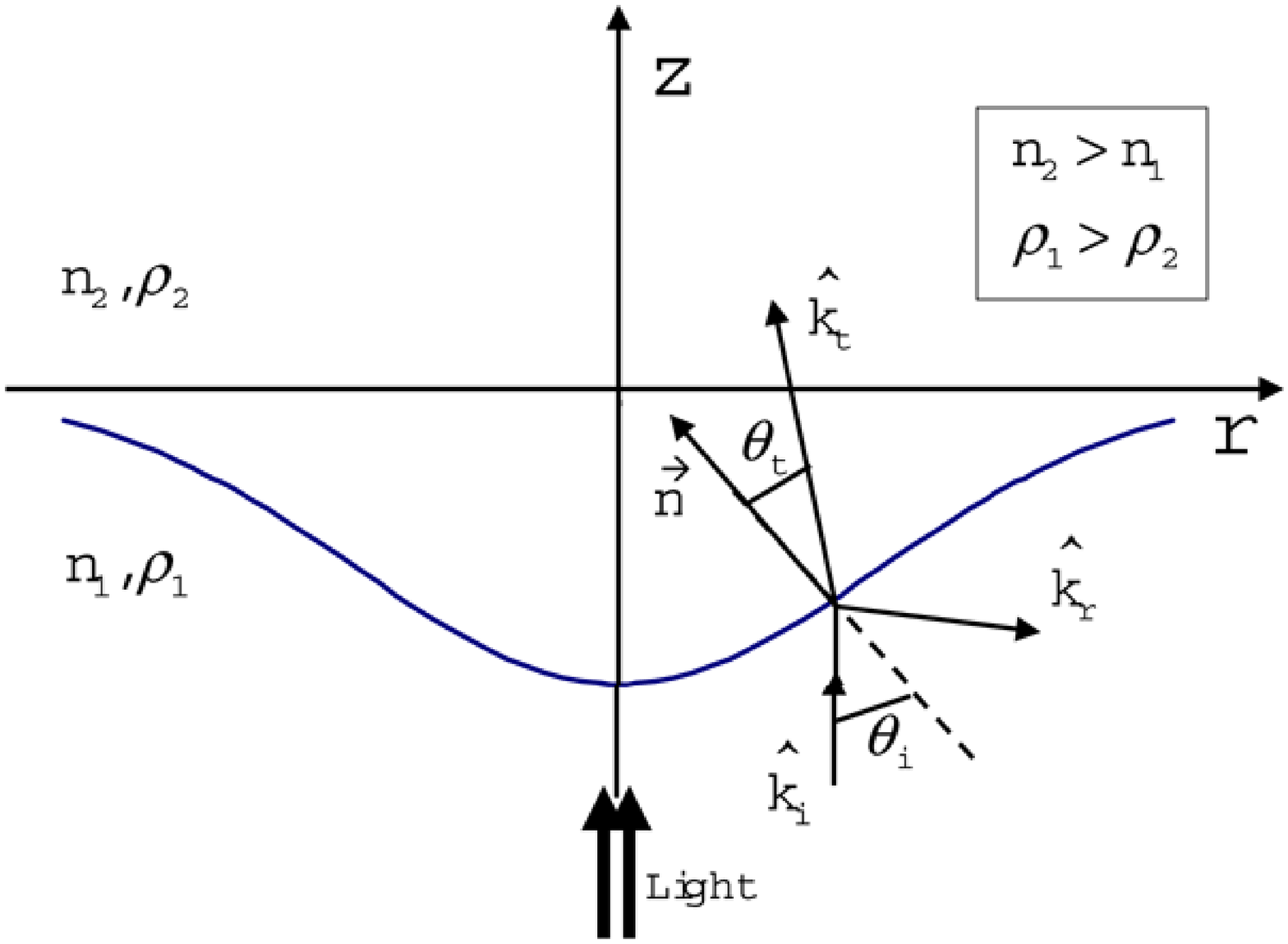}
\caption{Definition sketch of the displaced surface. The laser
illumination is from below.} \label{Fig1}
\end{figure}

\begin{figure}[h]
\centering
\includegraphics[height=8cm]{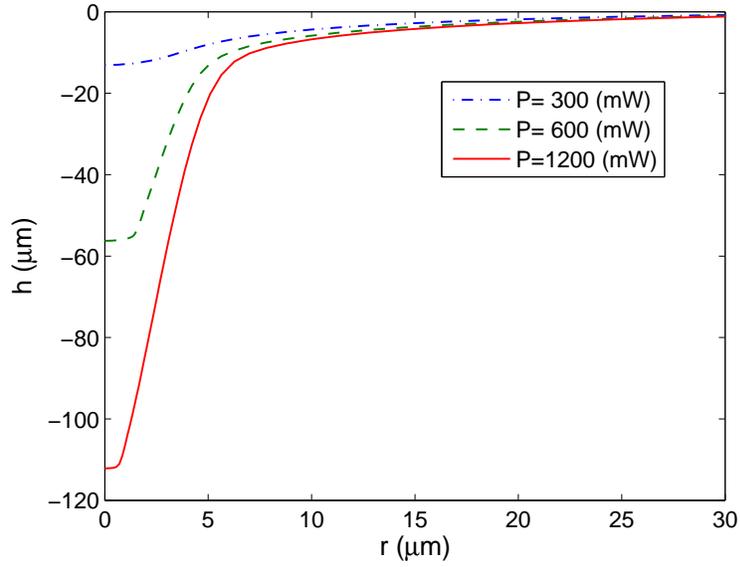}
\caption{Theoretical height $h(r)$ of displaced surface versus
radius $r$ when $T-T_C=2.5$ K, $\omega_0=4.8$ $\mu$m, for three
different laser powers $P$. The undisturbed surface is at
$h(r)=0$.} \label{Fig2}
\end{figure}

\eject
\begin{figure}[h]
\centering
\includegraphics[height=8cm]{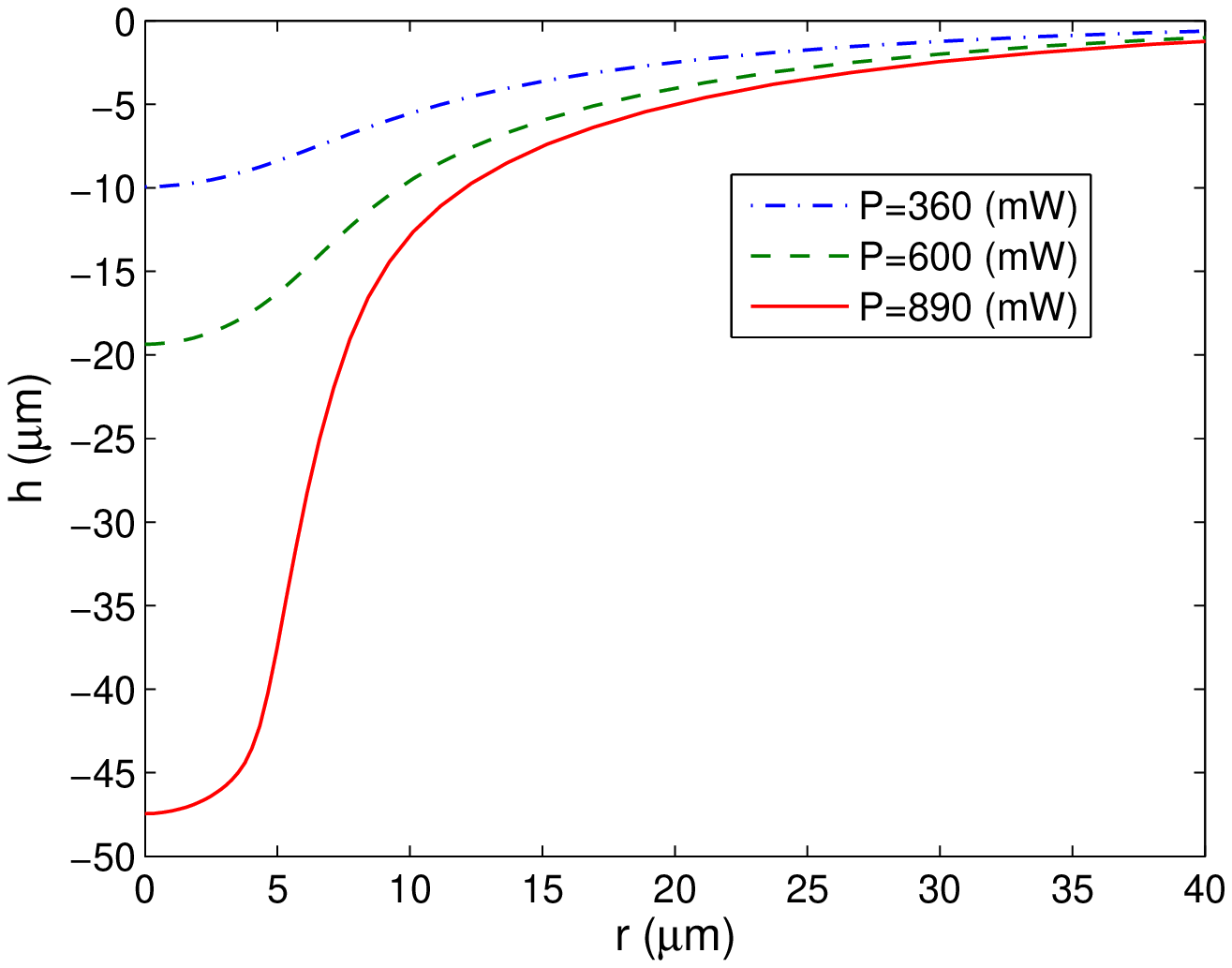}
\caption{Same as Fig.~2, but with $T-T_C=2.5$ K, $\omega_0=8.9$
$\mu$m.} \label{Fig3}
\end{figure}

\begin{figure}[h]
\centering
\includegraphics[height=8cm]{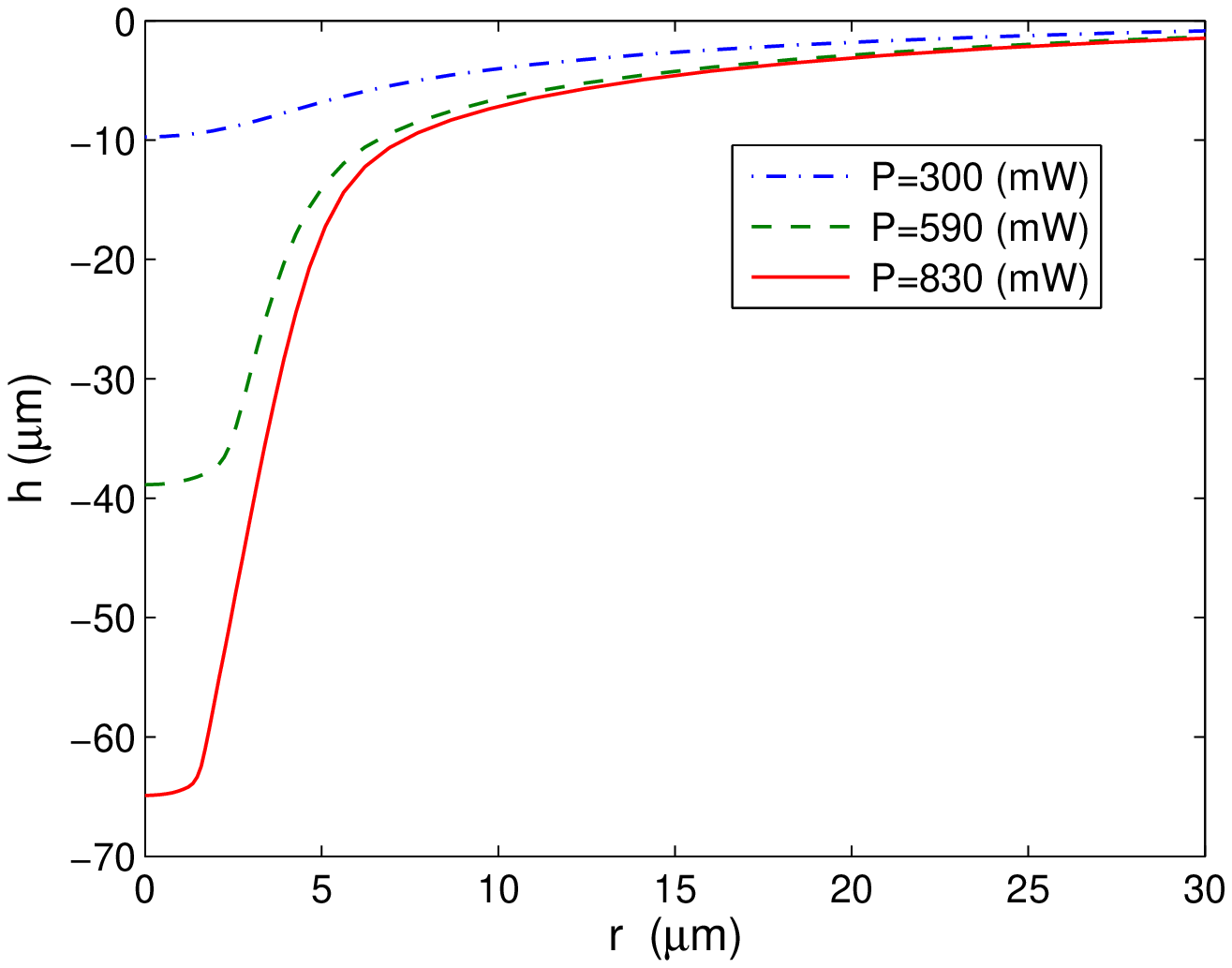}
\caption{Same as Fig.~2, but with $T-T_C=3$ K, $\omega_0=5.3$
$\mu$m.} \label{Fig4}
\end{figure}

\end{document}